\documentclass[aps]{revtex4}
\usepackage{amsmath}
\usepackage{amssymb,epsf}
\usepackage{latexsym}
\usepackage{color}

\begin{document}

 \title{Stability and Quasinormal Modes of Black Holes in Conformal Weyl Gravity}
 \author{Mehrab Momennia$^{1,2}$\footnote{email address: m.momennia@shirazu.ac.ir},
 Seyed Hossein Hendi$^{1,2,3}$\footnote{email address: hendi@shirazu.ac.ir} and
 Fatemeh Soltani Bidgoli$^{1,2}$}
 \affiliation{
 $^1$Department of Physics, School of Science, Shiraz University, Shiraz 71454, Iran \\
 $^2$Biruni Observatory, School of Science, Shiraz University, Shiraz 71454, Iran    \\
 $^3$ Canadian Quantum Research Center 204-3002 32 Ave Vernon, BC V1T 2L7 Canada
 }

\begin{abstract}
In this paper, we first investigate the thermal stability of black holes in
conformal Weyl gravity with a comparison with the Schwarzschild black holes.
Then, we consider a minimally coupled massive scalar perturbation and
calculate the quasinormal modes in asymptotically dS spacetime by employing
the sixth order WKB approximation and asymptotic iteration method. The
deviations from those of the Schwarzschild-dS solutions are obtained and the
possibility of the presence of quasi-resonance modes for Weyl black hole
solutions is investigated. Finally, we consider a massless scalar
perturbation in the background of asymptotically AdS solutions and calculate
the quasinormal modes by using the pseudospectral method. The effects of the
free parameter of the theory on the quasinormal modes are studied and
deviations from those of the Schwarzschild-AdS black holes are investigated.
The imaginary part of quasinormal frequencies in AdS spacetime is the time
scale of a thermal state (in the conformal field theory) to approach thermal
equilibrium.
\end{abstract}

\maketitle

\section{Introduction}

Considering the quantum effects in gravitational interaction, one may find
that the higher-curvature modification of general relativity is inevitable.
However, to have a physically ghost-free theory of higher-curvature
modifications, some special constraints should be applied. Fortunately,
there are known higher-curvature interesting renormalizable actions with no
ghosts under certain criteria. As an interesting example, we can regard the
so-called conformal gravity (CG), which its action is defined by the square
of the Weyl tensor \cite{Riegert,Lu}.

The CG is an interesting theory of modified general relativity with a
remarkable property that is sensitive to angles, but not distances. In other
words, it is invariant under local stretching of the metric which is called
the Weyl transformation, $g_{\mu \nu }(x)\rightarrow \Omega ^{2}(x)g_{\mu
\nu }(x)$. It has been shown that CG is useful for constructing supergravity
theories \cite{Bergshoeff} and can be considered as a possible UV completion
of general relativity \cite{Adler,Hooft,Mannheim}. It may also arise from
twister-string theory with both closed strings and gauge-singlet open
strings \cite{Berkovits}. Moreover, CG can appear as a counterterm in $%
AdS_{5}/CFT_{4}$ calculations \cite{Liu,Balasubramanian}. In addition to the
motivations mentioned above, solving the dark matter and dark energy
problems are two of the most important and interesting motivations of
studying CG theory (see \cite{Mannheim,MannheimPPNP} for more details).

Since CG is renormalizable \cite{Adler,Stelle} and the requirement of
conformal invariance at the classical level leads to a renormalizable gauge
theory of gravity, it will be interesting to consider black holes in CG
which permits a consistent picture of black hole evaporation \cite%
{Hasslacher}. The first attempt to obtain the spherically symmetric black
hole solutions in four dimensions has been done by Bach \cite{Bach}, and
then, Buchdahl has considered a particular case of the conformal solutions
in \cite{Buchdahl}. It is worthwhile to mention that the $4$-dimensional
solution of Einstein gravity is a solution of CG as well. Besides, it has
been shown that the Einstein solutions can be obtained by considering
suitable boundary conditions on the metric in CG \cite{Anastasiou}.

On the other hand, when a black hole undergoes perturbations, the resulting
behavior leads to some oscillations which are called quasinormal modes
(QNMs). The quasinormal frequencies (QNFs) related to such QNMs are
independent of initial perturbations and they are the intrinsic imprints of
the black hole response to external perturbations on the background
spacetime of black hole. The asymptotic behavior of the QNMs relates to the
quantum gravity \cite{Nollert} and the imaginary part of the frequencies in
AdS spacetime corresponds to the decay of perturbations of a thermal state
in the conformal field theory \cite{Horowitz,Lemos}.

The QNM is one of the most important and exciting features of compact
objects and describes the evolution of fields on the background spacetime of
such objects \cite{Kokkotas,KonoplyaR}. Therefore, the QNM spectrum reflects
the properties of spacetime, and consequently, we can find out about the
properties of background spacetime by studying the QNMs. As a result, the
QNM spectrum will be a function of black hole parameters, such as mass,
charge, and angular momentum. The QNM spectrum of gravitational
perturbations can be observed by gravitational wave detectors \cite{Abbott1}%
, and after the detection of the gravitational waves of compact binary
mergers by LIGO, investigation of the QNMs of black holes attracted
attention during the past three years (for instance, see an incomplete list
\cite{Konoplyajcap,Gonzalez,Kunz,Zhidenko,CardosoPRL} and references
therein). In the case of conformal gravity, the gravitational waves of small
perturbations in Minkowski background have been investigated and the
effective energy-momentum tensor of the gravitational radiation is
calculated \cite{Rongjia}. The astrophysical gravitational waves of
inspiralling compact binaries have been also investigated \cite{Caprini}.
More recently, the electromagnetic and gravitational perturbations of black
holes in conformal Weyl gravity have been studied \cite{MomenniaWeyl}, and
also, the QNMs of these black holes in the nearly extreme regime have been
obtained \cite{Momennia}. In this paper, we consider black hole solutions in
conformal Weyl gravity and investigate their thermal stability and QNMs of
scalar perturbations in asymptotically (A)dS spacetimes. We also investigate
deviations from the Schwarzschild-(A)dS black holes due to an additional
linear $r$-term in conformal Weyl solutions.

The outline of this paper is as follows. In the next section, we give a
brief review of $4$-dimensional black holes in conformal Weyl gravity. Then,
thermodynamic properties of the Weyl solutions are investigated and the
thermal stability analysis of the solutions is obtained. We also consider a
minimally coupled massive scalar perturbation in the background spacetime of
the black holes and calculate the related QNMs by using the sixth order WKB
approximation and the asymptotic iteration method (AIM). These calculations
are done in asymptotically dS spacetime, and the conditions of possible
quasi-resonance modes and anomalous decay rate of QNMs are studied. Finally,
the QNMs of the black holes in asymptotically AdS spacetime are obtained and
some relations for calculating the QNMs of large black holes are found. For
both dS and AdS cases, deviation of results from those of the
Schwarzschild-(A)dS black holes is discussed. We finish our paper with some
concluding remarks.

\section{review of solutions}

The four-dimensional conformal Weyl action reads
\begin{eqnarray}
I_{G} &=&-\alpha \int d^{4}x\sqrt{-g}C_{\lambda \mu \nu \kappa }C^{\lambda
\mu \nu \kappa }  \notag \\
&\equiv &-2\alpha \int d^{4}x\sqrt{-g}\left[ R^{\mu \nu }R_{\mu \nu }-{\frac{%
1}{3}}(R_{\phantom{\alpha}\alpha }^{\alpha })^{2}\right] ,  \label{IG}
\end{eqnarray}%
where the Weyl conformal tensor is
\begin{eqnarray}
C_{\lambda \mu \nu \kappa } &=&R_{\lambda \mu \nu \kappa }+{\frac{1}{6}}R_{%
\phantom{\alpha}\alpha }^{\alpha }\left[ g_{\lambda \nu }g_{\mu \kappa
}-g_{\lambda \kappa }g_{\mu \nu }\right]  \notag \\
&-&{\frac{1}{2}}\left[ g_{\lambda \nu }R_{\mu \kappa }-g_{\lambda \kappa
}R_{\mu \nu }-g_{\mu \nu }R_{\lambda \kappa }+g_{\mu \kappa }R_{\lambda \nu }%
\right] .  \label{C}
\end{eqnarray}

Variation of action (\ref{IG}) with respect to the metric tensor leads to
the following equation of motion
\begin{eqnarray}
&&W^{\mu \nu }=2C_{\ \ \ \ \ ;\lambda \kappa }^{\mu \lambda \nu \kappa
}-C^{\mu \lambda \nu \kappa }R_{\lambda \kappa }=  \notag \\
&&\frac{1}{2}g^{\mu \nu }(R_{\phantom{\alpha}\alpha }^{\alpha })_{%
\phantom{;\beta};\beta }^{;\beta }+R_{\phantom{\mu\nu;\beta};\beta }^{\mu
\nu ;\beta }-R_{\phantom{\mu\beta;\nu};\beta }^{\mu \beta ;\nu }-R_{%
\phantom{\nu \beta;\mu};\beta }^{\nu \beta ;\mu }-2R^{\mu \beta }R_{%
\phantom{\nu}\beta }^{\nu }  \notag \\
&&+\frac{1}{2}g^{\mu \nu }R_{\alpha \beta }R^{\alpha \beta }-\frac{2}{3}%
g^{\mu \nu }(R_{\phantom{\alpha}\alpha }^{\alpha })_{\phantom{;\beta};\beta
}^{;\beta }+\frac{2}{3}(R_{\phantom{\alpha}\alpha }^{\alpha })^{;\mu ;\nu }+%
\frac{2}{3}R_{\phantom{\alpha}\alpha }^{\alpha }R^{\mu \nu }-\frac{1}{6}%
g^{\mu \nu }(R_{\phantom{\alpha}\alpha }^{\alpha })^{2}=0.  \label{W}
\end{eqnarray}

\qquad It was shown that the static spherically symmetric solution of Weyl
gravity in four dimensions can be written as
\begin{equation}
ds^{2}=-f(r)dt^{2}+\frac{dr^{2}}{f(r)}+r^{2}d\Omega ^{2},  \label{metric}
\end{equation}%
where $d\Omega ^{2}$ is the line element of a $2-$sphere and the metric
function is \cite{Buchdahl}
\begin{equation}
f(r)=c+\frac{d}{r}+\frac{c^{2}-1}{3d}r+br^{2},  \label{f(r)}
\end{equation}%
where $b$, $c$, and $d$\ are integration constants. It is clear that for
non-vanishing $b$ that plays the role of the cosmological constant, Eq. (\ref%
{f(r)}) is not a vacuum solution of Einstein gravity with or without the
cosmological constant, while as long as $b=0$, $c=1$,\ and $d=-2M$, the
metric becomes identical to the Schwarzschild solution of Einstein gravity.
It is worth mentioning that although $b$ plays the role of the cosmological
constant, it is arisen purely as an integration constant and is not put in
the action by hand. Such a constant cannot be added to the action of WG
because it would introduce a length scale and hence breaks the conformal
invariance. For future comparison with the Schwarzschild-(A)dS black holes,
we set $b=-\Lambda /3\ $and$\ d=-2M$,\ and consider the following metric
function from now%
\begin{equation}
f(r)=c-\frac{2M}{r}-\frac{c^{2}-1}{6M}r-\frac{1}{3}\Lambda r^{2},  \label{MF}
\end{equation}%
where for $c=1$\ reduces to the Schwarzschild-(A)dS solutions. Fig. \ref%
{FofR} shows the behavior of the metric function (\ref{MF}) for different $c$%
. The singularity of Weyl solution can be covered with an event horizon, and
therefore, we can interpret the solution as a black hole.

\begin{figure}[tbp]
$%
\begin{array}{ccc}
\epsfxsize=7.5cm \epsffile{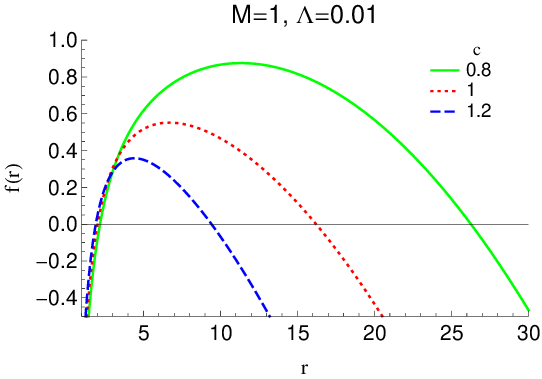} & \epsfxsize=7.5cm \epsffile{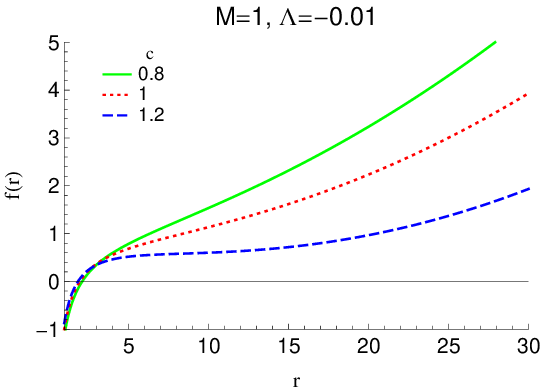} &
\end{array}
$%
\caption{The metric function versus radial coordinate. Positive $\Lambda$
corresponds to asymptotic dS solutions whereas the negative sign belongs to
AdS ones.}
\label{FofR}
\end{figure}

\section{thermodynamics \label{thermo}}

\subsection{Thermodynamic quantities}

Here, we calculate temperature and entropy of the Weyl black holes by using
the surface gravity at the event horizon and the Wald entropy formula,
respectively \cite{Wald}. After that, we investigate thermal stability of
the black holes.

By calculating the surface gravity, $\kappa =\sqrt{-\left( \nabla _{\mu
}\chi _{\nu }\right) \left( \nabla ^{\mu }\chi ^{\nu }\right) /2}$ ($\chi
=\partial _{t}$ is the null Killing vector of the horizon), we can obtain
the Hawking temperature of the black hole at the outermost (event) horizon, $%
r_{+}$. The temperature of the black hole is simplified as
\begin{equation}
T=\frac{\kappa }{2\pi }=\frac{1}{4\pi}\left. \frac{df(r)}{dr} \right\vert
_{r=r_{+}}=-\frac{1}{4\pi }\left( -\frac{2M_{+}}{r_{+}}+\frac{c^{2}-1}{6M_{+}%
}+\frac{2}{3}\Lambda r_{+}\right) ,  \label{temp}
\end{equation}%
in which $M_{+}$\ can be obtained by $f(r_{+})=0$ as%
\begin{equation}
M_{+}=\frac{r_{+}}{12}\left( 3c-\Lambda r_{+}^{2}\pm \sqrt{12-3c^{2}+\Lambda
^{2}r_{+}^{4}-6c\Lambda r_{+}^{2}}\right) ,  \label{mass}
\end{equation}%
and we consider the lower sign that is compatible with the
Schwarzschild-(A)dS black hole at $c\rightarrow 1$ limit. Thus, the
temperature converts to
\begin{equation}
T=-\frac{1}{12\pi r_{+}}\left( 2\Lambda r_{+}^{2}+\sqrt{12-3c^{2}+\Lambda
^{2}r_{+}^{4}-6c\Lambda r_{+}^{2}}\right) .  \label{TEMP}
\end{equation}

Besides, the entropy of the black hole in higher derivative theories can be
obtained by Wald's formula \cite{Wald,Iyer} which makes the dependence of
entropy on gravitational action%
\begin{equation}
S=-2\pi \alpha \int_{\mathcal{M}}d^{2}x\sqrt{h}\frac{\delta \mathcal{L}}{%
\delta R_{\mu \nu \rho \sigma }}\xi _{\mu \nu }\xi _{\rho \sigma },
\end{equation}%
where $\alpha $\ is a constant with dimension $\left( length\right) ^{2}$, $%
\mathcal{L}$\ is the Lagrangian density of the theory, $\xi _{\mu \nu }$\ is
the binormal to the (arbitrary) cross-section $\mathcal{M}$ of the horizon,
and $h$ is the determinant of induced metric on $\mathcal{M}$. Therefore,
the entropy of our case study black hole takes the following form
\begin{eqnarray}
S &=&-\frac{\alpha }{8}\int_{\mathcal{M}}d^{2}x\sqrt{h}C^{\mu \nu \rho
\sigma }\xi _{\mu \nu }\xi _{\rho \sigma }=\frac{6M_{+}+r_{+}\left(
1-c\right) }{6r_{+}}\omega \alpha  \notag \\
&=&\frac{\alpha \pi }{3}\left( 2+c-\Lambda r_{+}^{2}-\sqrt{12-3c^{2}+\Lambda
^{2}r_{+}^{4}-6c\Lambda r_{+}^{2}}\right) .  \label{entropy}
\end{eqnarray}

Now, we use the first law of thermodynamics ($\delta
M_{tot}=T\delta S$) to calculate the total mass of the black holes
as
\begin{equation}
M_{tot}=-\frac{2\alpha \Lambda M_{+}}{3}. \label{Mtot}
\end{equation}

Notably, for general $c$ ($c \neq 1$), one cannot obtain the
expression for the finite mass (\ref{Mtot}) by using either the
Deser-Tekin \cite{DT} or the AMD \cite{AMD1,AMD2} methods (see
\cite{Pope} for more details). Nevertheless, it is worth
mentioning that the validity of Eq. (\ref{Mtot}) is explicitly
checked in the appendix D of Ref. \cite{Pope} through the Noether
approach.

\subsection{Thermal stability}

Here, we explore thermal stability and the possibility of a phase
transition of the obtained black hole solutions. For an ordinary
thermodynamical system, thermal stability criteria can be governed
by the sign of heat capacity and compressibility. In other words,
the following conditions \cite{Callen}
\begin{equation}
c_{p}\geq c_{v}\geq 0\ \ \ \ \ \&\ \ \ \ \ \kappa _{T}\geq \kappa _{S}\geq 0,
\label{HeatCapacities}
\end{equation}%
guarantee the stability, and thus, heat capacities (at constant pressure $%
c_{p}$, or constant volume $c_{v}$) and compressibilities (isothermal $%
\kappa _{T}$, or isentropic $\kappa _{S}$) must be positive in a stable
system.

Now, we investigate the stability conditions of constructed black hole
solutions to find the thermally stable criteria. To do so, we should only
investigate a heat capacity since this system just has one intensive
parameter, $T$. The heat capacity of the solutions has the following
explicit form
\begin{eqnarray}
C &=&\partial _{T}M_{tot}=T\partial _{T}S=  \notag \\
&&-\frac{2\pi \Lambda r_{+}^{2}\left( \Lambda r_{+}^{2}-3c+\sqrt{%
12-3c^{2}+\Lambda ^{2}r_{+}^{4}-6c\Lambda r_{+}^{2}}\right) \left( 2\Lambda
r_{+}^{2}+\sqrt{12-3c^{2}+\Lambda ^{2}r_{+}^{4}-6c\Lambda r_{+}^{2}}\right)
}{9\left( c^{2}-4\right) +3\Lambda r_{+}^{2}\left( \Lambda r_{+}^{2}+2\sqrt{%
12-3c^{2}+\Lambda ^{2}r_{+}^{4}-6c\Lambda r_{+}^{2}}\right) },  \label{HC}
\end{eqnarray}%
where we set the constant $\alpha =1$ without loss of generality. The sign
of heat capacity help us to find the stability information. The positive
sign shows stable solutions whereas the negative sign indicates unstable
ones. The heat capacity changes sign whenever it meets root or divergence
points. The root of heat capacity indicates a bound point which separates
the physical black holes (positive temperature) from unstable ones (negative
temperature). In addition, divergence points may separate stable and
unstable regions and may signal the existence of phase transition.

\begin{figure}[tbp]
$%
\begin{array}{ccc}
\epsfxsize=7.5cm \epsffile{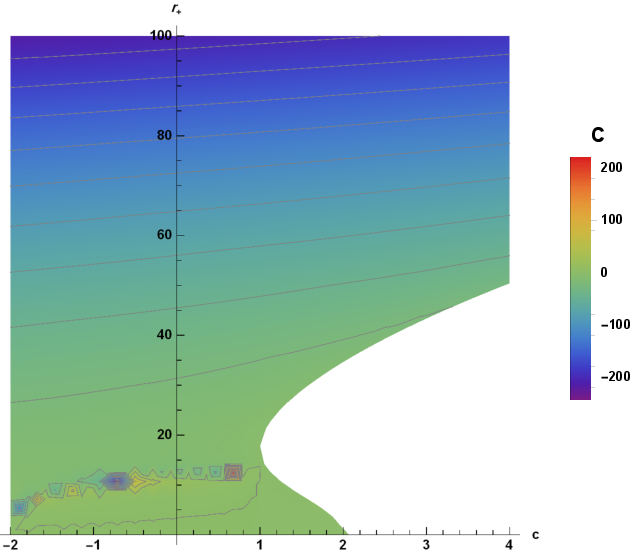} & \epsfxsize=7.5cm %
\epsffile{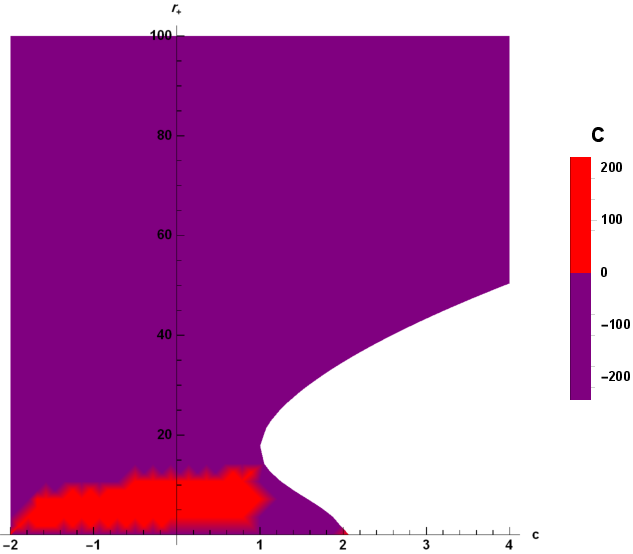} &
\end{array}
$%
\caption{The heat capacity versus $r_{+}$ and $c$ in
asymptotically dS spacetime for $\Lambda =0.01$ and $M=1$. The
white area is due to either divergency or imaginary/complex value
of $C$. The right panel is plotted for clarity of stable regions.}
\label{HCLP}
\end{figure}

\begin{figure}[tbp]
$%
\begin{array}{ccc}
\epsfxsize=7.5cm \epsffile{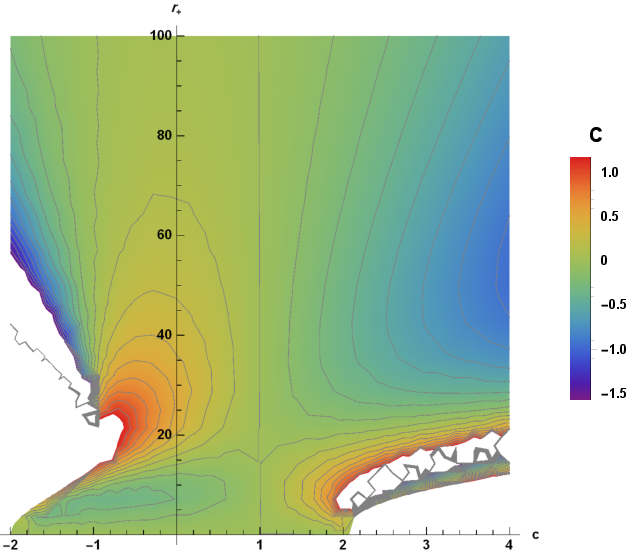} & \epsfxsize=7.5cm %
\epsffile{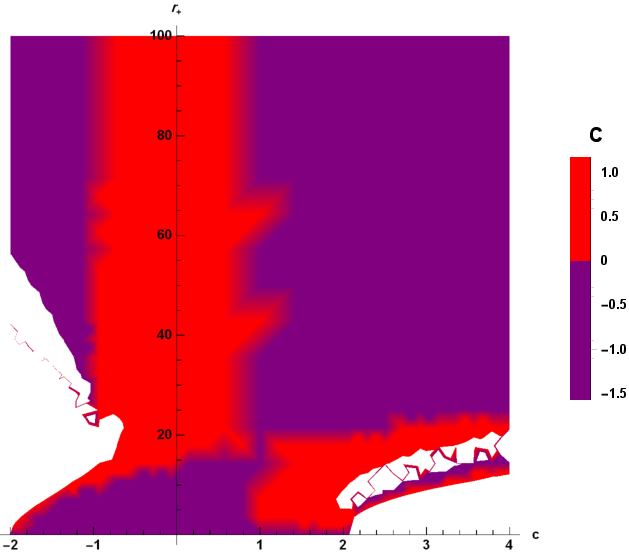} &
\end{array}
$%
\caption{The heat capacity versus $r_{+}$ and $c$ in
asymptotically AdS spacetime for $\Lambda =-0.01$ and $M=1$. The
white area is due to either divergency or imaginary/complex value
of $C$. The right panel is plotted for clarity of stable regions.}
\label{HCLM}
\end{figure}

Figures \ref{HCLP} and \ref{HCLM} show the behavior of the heat capacity
with different $c$ for small black holes (SBHs) and large black holes
(LBHs). We fixed the cosmological constant since we are going to investigate
the effect of $c$ which is a new parameter compared with the
Schwarzschild-(A)dS solutions. From these figures we find that the new
parameter $c$ affects the stability conditions. Besides, the behavior of the
heat capacity is different in the cases of SBHs and LBHs for $c\lesssim 1 $
and $c\gtrsim 1$. One may note that the SBHs of Einstein-$\Lambda$ gravity
(Schwarzschild-(A)dS solutions) are unstable for all values of $\Lambda$.
But in the case of Weyl solutions, there are stable SBHs for $c\lesssim 1$\ (%
$c\gtrsim 1$) and positive $\Lambda $ (negative $\Lambda $). In
contrast, the LBHs of Einstein-$\Lambda$ gravity are stable
everywhere. However, we have large stable black holes in Weyl
gravity just for negative $\Lambda $ and $-1\lesssim c\lesssim 1$.
Thus, the impact of new parameter $c$ is significant regarding the
stability conditions, especially for SBHs.

From this section, we found that the free parameter $c$ changes the
stability conditions of black holes in the Weyl gravity, significantly,
compared with the Schwarzschild-(A)dS black holes, i.e., stable black holes
become unstable and vice versa. Thus, we can expect to see interesting and
notable effects of $c$ on the QNMs of Weyl solutions, as we will show in the
following sections.


\section{quasinormal modes in dS spacetime \label{QNM}}

Scalar fields have been extensively investigated in the context of cosmology
as inflatons \cite{Cheung}, dark energy \cite{Gubitosi}, and dark matter
\cite{Hu}. They can be considered as candidates for constructing quantum
gravity \cite{Metsaev,Arvanitaki}, and also, can be considered as fields in
the strong field regime to modify the black hole background \cite%
{Herdeiro,Silva}. Scalar fields can be regarded to form clouds through
accretion or instabilities around black holes \cite{Brito,Clough}. In this
regard, one can consider non-minimal interaction between the scalar field
and gravity which leads to black hole hair \cite{Herdeiro,Babichev}. In
different models with non-minimal interaction of scalar fields with the
spacetime metric, the gravitational waves $h_{\mu \nu }$ of the metric will
be a linear combination of gravitational waves in general relativity, $%
h_{\mu \nu }^{(0)}$, and the scalar field solutions
\begin{equation}
h_{\mu \nu }=h_{\mu \nu }^{(0)}+\beta g_{\mu \nu }\Phi ,
\end{equation}%
where $\Phi $\ is the scalar field, $g_{\mu \nu }$\ is the
background metric, and $\beta $ is the non-minimal coupling
parameter. The QNMs $h_{\mu \nu }$, that could potentially be
observed, would have components oscillating with a combination of
the general relativity and scalar QNMs. Therefore, the signature
of scalar fields on gravitational waves could be observed by
future gravitational wave detectors. However, a minimally coupled
scalar field describes the QNMs in a range of scalar-tensor
theories. In this paper, we focus on minimal coupled scalar fields
to investigate the effects of the new parameter $c$ on the scalar
QNM spectrum in asymptotically dS spacetime and find deviations
from those of the Schwarzschild-dS black holes.

\subsection{Setup}

Here, we consider a massive scalar perturbation in the background of the
black hole spacetime and obtain the QNFs by using two independent methods of
calculations; the sixth order WKB approximation \cite%
{Schutz,IyerWill,Konoplya6th} and the asymptotic iteration method (AIM) \cite%
{AIM}. Furthermore, we concentrate our attention on the asymptotically dS
black holes ($\Lambda >0$) with the obtained metric function (\ref{MF}). The
asymptotic flat solutions ($c=1$ and $\Lambda =0$) reduce to the
Schwarzschild black holes which have been investigated extensively.

The equation of motion for a minimally coupled massive scalar field is given
by
\begin{equation}
\square \Phi -\mu ^{2}\Phi =0,  \label{WEQ}
\end{equation}%
in which $\mu $ is the mass of the scalar field $\Phi $ and $\square =\nabla
_{\nu }\nabla ^{\nu }$. Now, we consider the following expansion of modes
\begin{equation}
\Phi \left( t,r,\theta ,\varphi \right) =\sum_{l,m}\frac{1}{r}\Psi
_{l,m}\left( r\right) Y_{l,m}\left( \theta ,\varphi \right) e^{-i\omega t},
\label{sh}
\end{equation}%
where $Y_{l,m}\left( \theta ,\varphi \right) $ denotes the spherical
harmonics on $2$-sphere. Here and in the rest of this paper, we omit the
integral over frequency $\omega $ in the Fourier transform for notational
simplicity. Substituting the decomposition into Eq. (\ref{WEQ}), the
equation of motion (\ref{WEQ}) reduces to a wavelike equation for the radial
part $\Psi _{l}\left( r\right) $ in the following way
\begin{equation}
\left[ \partial _{r_{\ast }}^{2}+\omega ^{2}-V_{l}\left( r_{\ast }\right) %
\right] \Psi _{l}\left( r_{\ast }\right) =0.  \label{Weq}
\end{equation}

\begin{figure}[tbp]
$%
\begin{array}{ccc}
\epsfxsize=7.5cm \epsffile{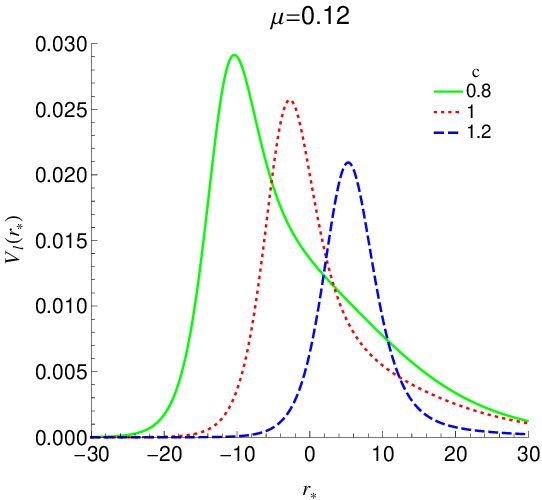} & \epsfxsize=7.5cm \epsffile{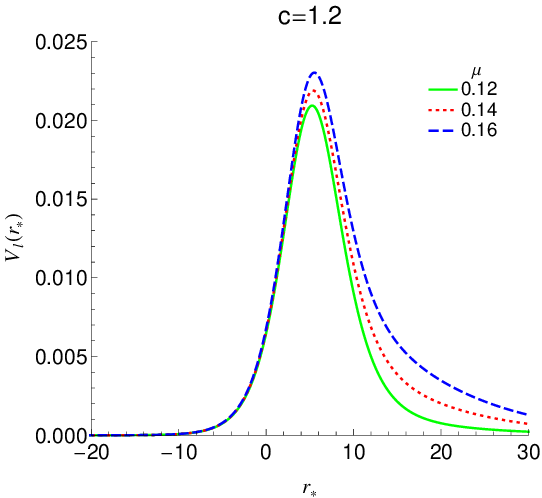} &
\end{array}
$%
\caption{Profiles of the effective potential versus tortoise coordinate for $%
M=1$, $\Lambda=0.01$, and $l=0$. The potential forms a barrier and vanishes
at both infinities.}
\label{Pot}
\end{figure}

In this equation, $r_{\ast }$ is the known tortoise coordinate with the
definition
\begin{equation}
dr_{\ast }=\frac{dr}{f(r)},  \label{tortoise}
\end{equation}%
and the effective potential $V_{l}\left( r_{\ast }\right) $ is given by
\begin{equation}
V_{l}\left( r_{\ast }\right) =f\left( r\right) \left[ \mu ^{2}+\frac{l\left(
l+1\right) }{r^{2}}+\frac{f^{\prime }\left( r\right) }{r}\right] ,
\label{EP}
\end{equation}%
where $l$ is the angular quantum number and it is notable that $r$
in
the right-hand side is a function of $r_{\ast }$ by (\ref{tortoise}). Figure %
\ref{Pot} shows the behavior of this effective potential (\ref{EP}) versus
the tortoise coordinate for different values of $c$ and $\mu $.

The spectrum of QNMs for a perturbed black hole spacetime is the solution of
the wave equation (\ref{Weq}). However, we have to impose some proper
boundary conditions to obtain its solutions. The quasinormal boundary
conditions imply that the wave at the event (cosmological) horizon is purely
incoming (outgoing)
\begin{equation}
\begin{array}{c}
\Psi _{l}\left( r\right) \sim e^{-i\omega r_{\ast }}\ \ \ \ \ \ as\ \ \ \ \
\ r_{\ast }\rightarrow -\infty \ (r\rightarrow r_{e}), \\
\Psi _{l}\left( r\right) \sim e^{i\omega r_{\ast }}\ \ \ \ \ \ \ as\ \ \ \ \
\ \ \ \ r_{\ast }\rightarrow \infty \ (r\rightarrow r_{c}),%
\end{array}
\label{bc}
\end{equation}%
where $r_{e}$ and $r_{c}$ are, respectively, the radius of event
horizon and the cosmological horizon. One should consider the
mentioned boundary conditions in order to obtain the QNFs.

\subsection{WKB approximation}

The method is based on the matching of WKB expansion of the wave function $%
\Psi _{l}\left( r_{\ast }\right) $ at the event horizon and
cosmological horizon with the Taylor expansion near the peak of
the potential barrier through two turning points. Therefore, this
method can be used for an effective potential that forms a
potential barrier and takes constant values (or zero) at the event
horizon ($r_{\ast }\rightarrow -\infty $) and cosmological horizon
($r_{\ast }\rightarrow \infty $) (like Fig. \ref{Pot}). The WKB
approximation was first applied to the problem of scattering
around black
holes \cite{Schutz}, and then extended to the third order \cite{IyerWill}, $%
6 $th order \cite{Konoplya6th} and recently to the $13$th order \cite{13th}.
The $6$th order of WKB formula is given by
\begin{equation}
\frac{i\left( \omega ^{2}-V_{0}\right) }{\sqrt{-2V_{0}^{\prime \prime }}}%
-\sum_{j=2}^{6}\Gamma _{j}=n+\frac{1}{2};\ \ \ \ \ \ n=0,1,2,...,
\end{equation}%
where $V_{0}$\ is the value of the effective potential at its local maximum,
the correction terms $\Gamma _{j}$'s\ correspond to the $j$th order and
depend on the value of the effective potential and its derivatives at the
local maximum, and $n$\ is the overtone number. The explicit form of the WKB
corrections is given in \cite{IyerWill}\ (for $\Gamma _{2}$ and $\Gamma _{3}$%
) and \cite{Konoplya6th} (for $\Gamma _{4}$, $\Gamma _{5}$, and $\Gamma _{6}$%
). It is worthwhile to mention that the WKB approximation does not give
reliable frequencies for $n\geq l$. We use this formula up to the sixth
order as a semi-analytical approach to obtain the QNFs of perturbations.

\subsection{AIM}

The AIM has been employed to solve the eigenvalue problems and solving
second-order differential equations \cite{Ciftci,CiftciHall}, and then it
was shown that it is an accurate technique for calculating QNMs \cite%
{Momennia,AIM,Naylor}. If one wants to employ the AIM, it is convenient to
use the independent variable $\xi =1/r$, and rewrite the wave equation (\ref%
{Weq}) as
\begin{equation}
\frac{d^{2}\Psi _{l}\left( \xi \right) }{d\xi ^{2}}+\frac{P^{\prime }}{P}%
\frac{d\Psi _{l}\left( \xi \right) }{d\xi }+\left( \frac{\omega ^{2}}{P^{2}}-%
\frac{V_{l}\left( \xi \right) }{P}\right) \Psi _{l}\left( \xi \right) =0,
\label{Wq2}
\end{equation}%
where $P$, $P^{\prime }$, and $V_{l}\left( \xi \right) $ are given by
\begin{eqnarray}
P &=&\xi ^{2}f\left( \xi \right) , \\
P^{\prime } &=&\frac{dP}{d\xi }=2c\xi -6M\xi ^{2}-\frac{c^{2}-1}{6M}, \\
V_{l}\left( \xi \right) &=&\left[ \frac{\mu ^{2}}{\xi ^{2}}+l\left(
l+1\right) +\frac{f^{\prime }\left( \xi \right) }{\xi }\right] , \\
f^{\prime }\left( \xi \right) &=&2M\xi ^{2}-\frac{c^{2}-1}{6M}-\frac{%
2\Lambda }{3\xi }.
\end{eqnarray}

In order to choose the appropriate scaling behavior for quasinormal boundary
conditions, one may define \cite{Moss,Naylor}
\begin{equation}
e^{i\omega r_{\ast }}=\prod_{j}\left( \xi -\xi _{j}\right) ^{i\omega /\kappa
_{j}},  \label{scaling}
\end{equation}%
in which $\kappa _{j}$ is the surface gravity at $\xi _{j}$ with $f\left(
\xi =\xi _{j}\right) =0$. This equation scale out the divergent behavior at
some boundary $\xi _{j}$ and applies the boundary conditions (\ref{bc}) to
the solution.

Now, based on Eqs. (\ref{bc}) and (\ref{scaling}), an appropriate choice for
QNMs to scale out the divergent behaviour at the cosmological horizon is as
follows
\begin{equation}
\Psi _{l}\left( \xi \right) =e^{i\omega r_{\ast }}\psi _{l}\left( \xi
\right) ,  \label{redifine}
\end{equation}%
so that the equation (\ref{Wq2}) converts to
\begin{equation}
P\frac{d^{2}\psi _{l}\left( \xi \right) }{d\xi ^{2}}+\left( P^{\prime
}-2i\omega \right) \frac{d\psi _{l}\left( \xi \right) }{d\xi }-V_{l}\left(
\xi \right) \psi _{l}\left( \xi \right) =0.  \label{Weq3}
\end{equation}

Besides, by considering the equations (\ref{scaling}) and (\ref{redifine}),
the correct quasinormal condition at the event horizon, $\xi _{e}$, is
\begin{equation}
\psi _{l}\left( \xi \right) =\left( \xi -\xi _{e}\right) ^{-i\omega /\kappa
_{e}}\mathcal{U}_{l}\left( \xi \right) ,  \label{SC}
\end{equation}%
where $\mathcal{U}_{l}\left( \xi \right) $ should be a finite and convergent
function, and $\kappa _{e}$\ is the surface gravity at the horizon
\begin{equation}
\kappa _{e}=\left. \frac{1}{2}\frac{df\left( r\right) }{dr}\right\vert
_{r=r_{e}}=\frac{1}{2}f^{\prime }\left( \xi _{e}\right) .
\end{equation}

By inserting (\ref{SC}) into (\ref{Weq3}), one can find the standard AIM
form as follows
\begin{equation}
\frac{d^{2}\mathcal{U}_{l}\left( \xi \right) }{d\xi ^{2}}=\lambda _{0}\left(
\xi \right) \frac{d\mathcal{U}_{l}\left( \xi \right) }{d\xi }+s_{0}\left(
\xi \right) \mathcal{U}_{l}\left( \xi \right) ,  \label{Weq4}
\end{equation}%
so that $\lambda _{0}\left( \xi \right) $\ and $s_{0}\left( \xi \right) $\
are
\begin{eqnarray}
\lambda _{0}\left( \xi \right) &=&-\frac{1}{P}\left( P^{\prime }-2i\omega -%
\frac{2i\omega P}{\kappa _{e}\left( \xi -\xi _{e}\right) }\right) ,
\label{l0} \\
s_{0}\left( \xi \right) &=&\frac{1}{P}\left[ \frac{i\omega \left( P^{\prime
}-2i\omega \right) }{\kappa _{e}\left( \xi -\xi _{e}\right) }-\frac{i\omega P%
}{\kappa _{e}\left( \xi -\xi _{e}\right) ^{2}}\left( \frac{i\omega }{\kappa
_{e}}+1\right) +V_{l}\left( \xi \right) \right] .  \label{s0}
\end{eqnarray}

Once the standard AIM form of the master wave equation is obtained, we can
differentiate it and apply the \textit{quantization condition} to calculate
the QN frequencies (see \cite{AIM,Naylor} for details of calculations).

\subsection{Results and discussion}

The QNMs are calculated by using the sixth order WKB approximation and AIM
after $15$ iterations, and results are presented in tables $I$-$IV$. The
tables contain the fundamental QNM ($n=0$) for different values of the new
parameter $c$\ and multipole number for fixed $M=1$, $\Lambda =0.01$, and $%
\mu =0.12$.

From table $I$, we find that although the WKB approximation does not give
reliable frequencies for $n\geq l$, for higher $c$, say $c\geq 1.15$, the
WKB formula gives better results for $n=0=l$. Also, most of the results of
WKB approximation are in a good agreement with numeric results (tables $II-IV
$), and results get better for the higher multipole number, $l>n$, for all
values of $c$ as we expected. On the other hand, both the real and imaginary
parts of the QNFs decrease with increasing $c$ which shows that there are
fewer oscillations for higher $c$ at the ringdown stage and the modes decay
faster for lower $c$. The tables $I$-$IV$ also show deviations of the QNMs
of Weyl solutions from those of the Schwarzschild-dS black holes. For lower $%
c$ ($c<1$), there are more oscillations and long life modes compared to
higher values of $c$ ($c>1$). Thus, the QNMs spectrum of Weyl solutions
deviates from those of the Schwarzschild-dS black holes and these deviations
could potentially be observed by using future gravitational wave detectors
for non-minimal interactions of scalar fields with the spacetime metric.

On the other hand, as one can see from Fig. \ref{Pot} that the effective
potential forms a potential barrier which is positive everywhere and
vanishes at the event horizon and spatial infinity. This shows that we can
find dynamically stable black holes undergoing massive scalar perturbations.
It is worthwhile to mention that although the conformal Weyl solutions are
dynamically stable for $c>1$, they are thermally unstable (table $I$). In
order to find black holes that are both thermally and dynamically stable for
$c>1$, one can consider larger black holes with $r_{+}>15$\ (see the right
panel of Fig. \ref{HCLP}).

\begin{center}
\begin{tabular}{|c|c|c|c|c|c|}
\hline\hline
$c$ &  & AIM ($\omega _{r}-i\omega _{i}$) & WKB ($\omega _{r}-i\omega _{i}$)
& $r_{+}$ & Thermal Stability \\ \hline\hline
$0.80$ &  & $0.1366-0.05443i$ & $0.1204-0.06786i~\left(
11.86\%,24.67\%\right) $ & $2.1853$ & Stable \\ \hline
$0.85$ &  & $0.1291-0.05818i$ & $0.1165-0.07347i~\left(
9.76\%,26.28\%\right) $ & $2.1419$ & Stable \\ \hline
$0.90$ &  & $0.1212-0.06243i$ & $0.1130-0.07898i~\left(
6.77\%,26.51\%\right) $ & $2.1012$ & Stable \\ \hline
$0.95$ &  & $0.1129-0.06735i$ & $0.1097-0.08424i~\left(
2.83\%,25.08\%\right) $ & $2.0633$ & Stable \\ \hline
$1$ &  & $0.1043-0.07325i$ & $0.1063-0.08915i~\left( 1.92\%,21.71\%\right) $
& $2.0278$ & Transition Point \\ \hline
$1.05$ &  & $0.09540-0.08087i$ & $0.1029-0.09368i~\left(
7.86\%,15.84\%\right) $ & $1.9947$ & unstable \\ \hline
$1.10$ &  & $0.08835-0.09201i$ & $0.09942-0.09776i~\left(
12.53\%,6.25\%\right) $ & $1.9639$ & unstable \\ \hline
$1.15$ &  & $0.08862-0.1014i$ & $0.09578-0.1014i~\left(
8.08\%,<0.01\%\right) $ & $1.9352$ & unstable \\ \hline
$1.20$ &  & $0.08799-0.1051i$ & $0.09198-0.1045i~\left( 4.53\%,0.57\%\right)
$ & $1.9086$ & unstable \\ \hline\hline
\end{tabular}%
\vspace{0.2cm}

Table $I$: The fundamental QNM for $l=0$. \\[0pt]
The special case of $c=1$ corresponds to the QNFs of the Schwarzschild-dS
black holes.\\[0pt]
\end{center}

\vspace{0.1cm}

\begin{center}
\begin{tabular}{|c|c|c|c|}
\hline\hline
$l$ &  & AIM ($\omega _{r}-i\omega _{i}$) & WKB ($\omega _{r}-i\omega _{i}$)
\\ \hline\hline
$0$ &  & $0.1366-0.05443i$ & $0.1204-0.06786i\ \left( 11.86\%,24.67\%\right)
$ \\ \hline
$1$ &  & $0.2867-0.08647i$ & $0.2871-0.08644i\ \left( 0.14\%,0.03\%\right) $
\\ \hline
$2$ &  & $0.4621-0.08879i$ & $0.4621-0.08878i\ \left( <0.01\%,0.01\%\right) $
\\ \hline
$3$ &  & $0.6401-0.08953i$ & $0.6401-0.08953i\ \left( <0.01\%,<0.01\%\right)
$ \\ \hline\hline
\end{tabular}%
\vspace{0.2cm}

Table $II$: The fundamental QNM for $c=0.8$.\\[0pt]
\end{center}

\vspace{0.1cm}

\begin{center}
\begin{tabular}{|c|c|c|c|}
\hline\hline
$l$ &  & AIM ($\omega _{r}-i\omega _{i}$) & WKB ($\omega _{r}-i\omega _{i}$)
\\ \hline\hline
$0$ &  & $0.1043-0.07325i$ & $0.1063-0.08915i\ \left( 1.92\%,21.71\%\right) $
\\ \hline
$1$ &  & $0.2833-0.09103i$ & $0.2834-0.09108i\ \left( 0.04\%,0.05\%\right) $
\\ \hline
$2$ &  & $0.4641-0.09147i$ & $0.4641-0.09147i\ \left( <0.01\%,<0.01\%\right)
$ \\ \hline
$3$ &  & $0.6463-0.09162i$ & $0.6463-0.09162i\ \left( <0.01\%,<0.01\%\right)
$ \\ \hline\hline
\end{tabular}%
\vspace{0.2cm}

Table $III$: The fundamental QNM for $c=1$. \\[0pt]
This table indicates the QNFs of the Schwarzschild-dS black holes.\\[0pt]
\end{center}

\vspace{0.1cm}

\begin{center}
\begin{tabular}{|c|c|c|c|}
\hline\hline
$l$ &  & AIM ($\omega _{r}-i\omega _{i}$) & WKB ($\omega _{r}-i\omega _{i}$)
\\ \hline\hline
$0$ &  & $0.08799-0.1051i$ & $0.09198-0.1045i\ \left( 4.53\%,0.57\%\right) $
\\ \hline
$1$ &  & $0.2704-0.09184i$ & $0.2704-0.09190i\ \left( <0.01\%,0.07\%\right) $
\\ \hline
$2$ &  & $0.4508-0.09076i$ & $0.4508-0.09077i\ \left( <0.01\%,0.01\%\right) $
\\ \hline
$3$ &  & $0.6312-0.09047i$ & $0.6312-0.09047i\ \left( <0.01\%,<0.01\%\right)
$ \\ \hline\hline
\end{tabular}%
\vspace{0.2cm}

Table $IV$: The fundamental QNM for $c=1.2$.
\end{center}

Besides, one of the motivations for considering the test massive fields
comes from the fact that there are some QNMs with arbitrarily long life
(purely real) modes called quasi-resonance modes \cite{Ohashi}. For the
quasi-resonance modes, the oscillations do not decay and the situation is
similar to the standing waves on a string that is fixed at its both ends.
The quasi-resonance modes occur for special values of the field mass and the
QNMs disappear when the field mass takes higher values. However, this
happens just for lower overtones whenever the effective potential is
non-zero at least at one of the boundaries (the event horizon $r_{\ast
}\rightarrow -\infty $\ or cosmological horizon $r_{\ast }\rightarrow \infty
$).

\begin{figure}[tbp]
$%
\begin{array}{ccc}
\epsfxsize=7.5cm \epsffile{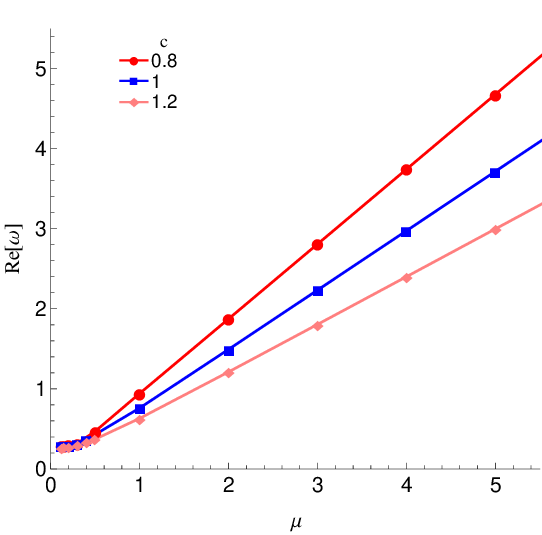} & \epsfxsize=7.5cm \epsffile{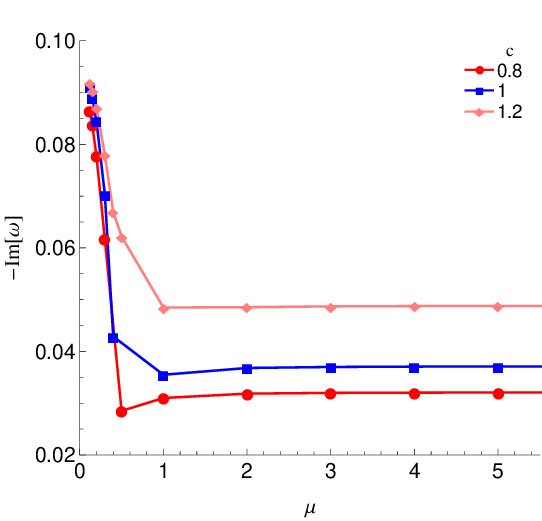} &
\end{array}
$%
\caption{The real and imaginary parts of the fundamental overtone as a
function of scalar field mass calculated by using the WKB formula for $M=1$,
$\Lambda =0.01$, and $l=1$.}
\label{muFig}
\end{figure}

Now, let us investigate the possibility of the quasi-resonance modes for
Weyl black hole solutions. The effective potential (\ref{EP})\ vanishes at
both infinities for all possible values of different parameters, and
therefore, there are no quasi-resonance modes for (\ref{Weq}). In addition,
if one sets the integration constant $\Lambda$ equals to zero, the effective
potential still vanishes at both infinities. There is only one case so that
the effective potential can be non-zero at spatial infinity and that is zero
cosmological constant ($\Lambda =0$) with $c=1$. In this case, the effective
potential reduces to the Schwarzschild black holes in flat spacetime which
its quasi-resonance modes have been investigated before \cite{SchwMSF}.
Therefore, our black hole case study has no quasi-resonant oscillations in
general and the imaginary part of the frequencies never vanishes (regardless
of the mentioned trivial case). Figure \ref{muFig} shows the behavior of
QNFs with increasing in $\mu $ and confirms the above discussion. As $\mu$
increases, the real part of frequencies increases too, whereas the imaginary
part first decreases rapidly and then takes a constant value.

\subsection{Anomalous decay rate of QNMs in dS spacetime}

It was recently shown that the decay timescales of the QNMs of a massive
scalar field perturbations in the Schwarzschild background either grow or
decay with increasing multipole number $l$, depending on whether the mass of
the scalar field is small or large \cite{Lagos}. This anomalous behavior is
due to an additional $\mu ^{2}$-term in the sub-leading term in the eikonal
expression for $\omega _{i}$. In this scenario, there is a critical mass $%
\tilde{\mu}$\ so that the imaginary part of the QNFs increases (decreases)
with increasing in $l$ for $\mu >\tilde{\mu}$\ ($\mu <\tilde{\mu} $). For
low-$l$ values, the value of the critical mass $\tilde{\mu}$ decreases when $%
l$\ increases, but there is one fixed critical mass for large-$l$ values.
\begin{figure}[tbp]
$%
\begin{array}{cccc}
\epsfxsize=5cm \epsffile{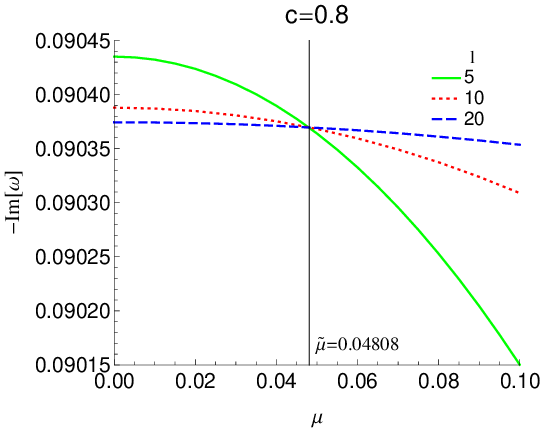} & \epsfxsize=5cm \epsffile{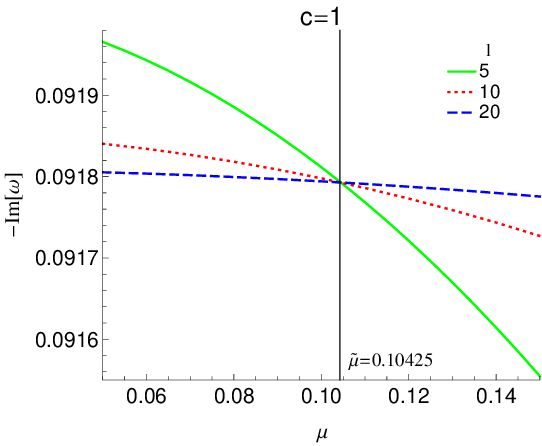} & %
\epsfxsize=5cm \epsffile{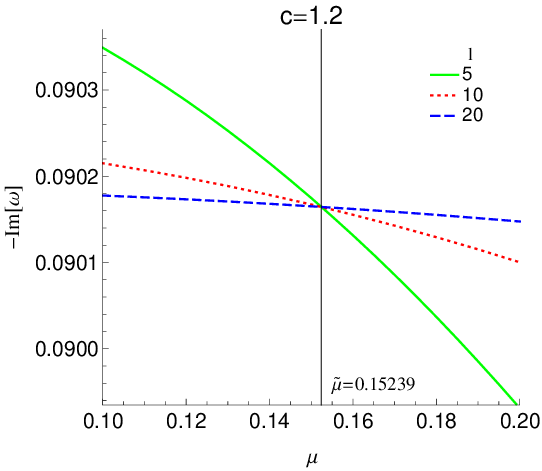} &
\end{array}
$%
\caption{The imaginary part of the fundamental overtone as function of $%
\protect\mu $ calculated by using the WKB formula for $M=1$ and $\Lambda
=0.01$. The vertical black line indicates the critical mass $\tilde{\protect%
\mu}$ where the curves cross each other for large-$l$ values ($c=1$
corresponds to the Schwarzschild black holes in dS spacetime).}
\label{Anomalous}
\end{figure}

Here, we investigate the possibility of this anomalous behavior for the Weyl
solutions and Schwarzschild black holes in dS spacetime numerically. Fig. %
\ref{Anomalous} shows the numerical results for $\omega _{i}$ as a function
of $\mu $ for different values of $l$\ and $c$. From the middle panel of
this figure, we find that the curves cross over at a special mass, and thus
the QNM spectrum of Schwarzschild black holes in dS spacetime contains this
anomaly as for the flat case \cite{Lagos}. Also, one can see the same
behavior for the Weyl solutions (left and right panels of Fig. \ref%
{Anomalous}). It is worthwhile to note that the free parameter $c$\ affects
the value of the critical mass and $\tilde{\mu}$\ increases with increasing
in $c$.

\section{quasinormal modes in AdS spacetime}

Here, we want to point out an application of the conformal Weyl black hole
solutions in the context of AdS/CFT correspondence. The AdS/CFT
correspondence describes a relation between the string theory on
asymptotically AdS spacetime and conformal field theory on the boundary \cite%
{Maldacena}. Some phenomena like the Nernst effect \cite{Hartnoll}, Hall
effect \cite{Kovtun}, superconductivity \cite{HartnollHerzog}, and the
decaying time scale of perturbations of a thermal state \cite{Horowitz} in
the field theory have dual descriptions in gravitational theory. Besides,
this correspondence between a gravitational theory and a CFT can be extended
to explain some aspects of nuclear physics \cite{Mateos}.

In terms of the AdS/CFT correspondence, a large AdS black hole corresponds
to an approximately thermal state in conformal field theory and scalar
perturbations of the black hole correspond to perturbations of this thermal
state. Therefore, the decay rate of the scalar field perturbations describes
the decay of perturbations of this thermal state. In this scenario, we can
calculate the QNMs of a large static black hole in asymptotically AdS
spacetime to obtain the time scale of the thermal state to approach thermal
equilibrium. Here, we shall obtain the QNMs of Weyl solutions with AdS
asymptote to find the stability time scale of the corresponding thermal
state. Investigating the dynamical stability of Weyl solutions undergoing
scalar perturbations is another advantage of calculating the QNMs. We can
follow either Horowitz-Hubeny approach \cite{Horowitz} or pseudospectral
method \cite{Boyd} to calculate the QNMs of asymptotically AdS black holes ($%
\Lambda <0$). In this paper, we follow the pseudospectral method and use a
public code given in \cite{Jansen} to obtain the QNMs. The pseudospectral
method replaces the continuous variable by a discrete set of points and
solves the resulting eigenvalue equation.

Now, we consider the fluctuations of a massless scalar field in the
background of Weyl solutions. It is convenient to obtain the master wave
equation in Eddington-Finkelstein coordinates whenever we are going to use
the pseudospectral method. In Eddington-Finkelstein coordinates, the
background line element (\ref{MF}) is given by
\begin{eqnarray}
ds^{2} &=&-f(u)dt^{2}-2u^{-2}dtdu+u^{-2}\left( d\theta ^{2}+\sin ^{2}\theta
d\varphi ^{2}\right) , \label{NMF1} \\
f(u) &=&c-2Mu-\frac{c^{2}-1}{6Mu}+\frac{1}{u^{2}L^{2}},
\label{NMF}
\end{eqnarray}%
where $u=1/r$\ and $L$\ is the AdS radius related to the cosmological
constant by $\Lambda =-3/L^{2}$. Thus, $u=0$\ represents the boundary and $%
u=1/r_{+}$\ corresponds to the event horizon. The equation of motion for a
minimally coupled massless scalar field is governed by Eq. (\ref{WEQ}) with $%
\mu =0$. By considering (\ref{NMF1}) as the spacetime background
and expanding the scalar field eigenfunction $\Phi $\ versus the
spherical harmonics as (\ref{sh}) and substituting the scalar
field decomposition (\ref{sh}) into (\ref{WEQ}), we can find the
following
second-order differential equation for the radial part%
\begin{equation}
u^{3}f(u)\Psi ^{\prime \prime }\left( u\right) +\left[ 2i\omega
u+u^{3}f^{\prime }\left( u\right) \right] \Psi ^{\prime }\left( u\right) -%
\left[ 2i\omega +ul\left( l+1\right) \right] \Psi \left( u\right) =0
\label{DEq}
\end{equation}%
in which $l$\ is the multipole number and $\omega =\omega _{r}-i\omega _{i}$%
\ is the quasinormal frequency with an imaginary part $\omega _{i}$\ giving
the damping of the perturbations and a real part $\omega _{r}$\ giving the
actual oscillations. Thus, in terms of the AdS/CFT correspondence, $\tau
=1/\omega _{i}$\ is the time scale that the thermal state should pass to
meet the thermal equilibrium. On the other hand, the negativity of the
imaginary part of the frequencies guarantees the dynamical stability of the
black hole \cite{Horowitz}. Otherwise, the perturbations increase as time
increases and the spacetime becomes unstable.

For the boundary conditions in AdS spacetime, causality requires ingoing
modes at the event horizon and finite modes at the boundary that results in
a discrete spectrum of QNFs $\omega $. To analyze the behavior of modes $%
\Psi \left( u\right) $ near the horizon and the boundary to see how to deal
with the boundary conditions, without loss of generality, we consider $%
r_{+}=1$ and replace the value of $M$ from the lower sign of Eq. (\ref{mass}%
). Therefore, $u=0$ represents the spatial infinity and $u=1$ corresponds to
the event horizon radius. Starting with the horizon, by substituting an
ansatz $\Psi \left( u\right) =\left( 1-u\right) ^{p}$ in Eq. (\ref{DEq}),
one can obtain two solutions $\Psi _{in}\left( u\right) \propto Const. +%
\mathcal{O}\left( 1-u\right) $\ and $\Psi _{out}\left( u\right) \propto
\left( 1-u\right) ^{i\omega \lambda }\left[ 1+\mathcal{O}\left( 1-u\right) %
\right] $\ where $\lambda $\ is a constant with the following explicit form
\begin{equation}
\lambda =\frac{6L^{2}}{6+\sqrt{9+18cL^{2}-3L^{4}\left( c^{2}-4\right) }}.
\end{equation}

By considering the time dependence $e^{-i\omega t}$ of the modes, the $\Psi
_{out}\left( t,u\right) $\ behaves as%
\begin{equation}
\Psi _{out}\left( t,u\right) \propto e^{-i\omega \left[ t-\lambda \ln \left(
1-u\right) \right] }.
\end{equation}

In order to keep a constant phase, $\left( 1-u\right) $\ has to increase as $%
t$\ increases, and thus $u$\ should decrease which means that this solution
is outgoing. Therefore, we must consider just the ingoing solution $\Psi
_{in}\left( u\right) \propto Const$ and discard the outgoing one. In
addition, the ingoing mode is perfectly smooth near the horizon while as we
approach the horizon, the outgoing mode oscillates more and more rapidly.
Fortunately, the boundary condition of ingoing modes at the horizon is
enforced automatically since we are working in the ingoing
Eddington-Finkelstein coordinates.

On the other hand, there are two solutions near the boundary; a normalizable
mode $\Psi \left( u\right) \propto u^{3}$\ and a non-normalizable one $\Psi
\left( u\right) \propto Const$ which we must discard. If we redefine $\Psi
\left( u\right) =u^{2}\tilde{\Psi}\left( u\right) $, then the normalizable
mode tends to zero linearly whereas the non-normalizable mode diverges as $%
\sim u^{-2}$. Doing this rescaling, the wave equation (\ref{DEq}) converts
to
\begin{equation}
u^{2}\mathcal{X}\tilde{\Psi}^{\prime \prime }\left( u\right) +u\left( 4%
\mathcal{X}+u\mathcal{Y}\right) \tilde{\Psi}^{\prime }\left( u\right) +\left[
2\mathcal{X}+2u\mathcal{Y}+u^{2}\left( ul\left( l+1\right) +2i\omega \right) %
\right] \tilde{\Psi}\left( u\right) =0,  \label{deq}
\end{equation}%
where%
\begin{eqnarray}
\mathcal{X} &=&-\frac{1}{L^{2}}u+\frac{c^{2}-1}{6M_{+}}%
u^{2}-cu^{3}+2M_{+}u^{4}, \\
\mathcal{Y} &=&\frac{2}{L^{2}}+\frac{1-c^{2}+12M_{+}\left(
M_{+}u^{2}-i\omega \right) }{6M_{+}}u,
\end{eqnarray}%
in which $M_{+}$ is given in (\ref{mass}). Now, the normalizable solution
behaves smoothly at the boundary and it should be considered, while we
discard the other non-normalizable solution. The wave equation (\ref{deq})
is an input for the code and one can fix the free parameters $L$ and $c$,
and also, the event horizon radius $r_{+}$ or $u_{+}$ (presented in $M_{+}$)
to calculate the QNMs.

We recall that the large black holes correspond to the thermal states in
CFT. Therefore, we shall focus on the QNMs of LBHs ($r_{+}>>L$) and discard
the small ones ($r_{+}<<L$). In addition, we set $L=1$\ as the AdS radius to
compare the Weyl solutions in asymptotically AdS spacetime with the
Schwarzschild-AdS black holes investigated in \cite{Horowitz,SchwADS}.

\begin{figure}[tbp]
$%
\begin{array}{ccc}
\epsfxsize=7.5cm \epsffile{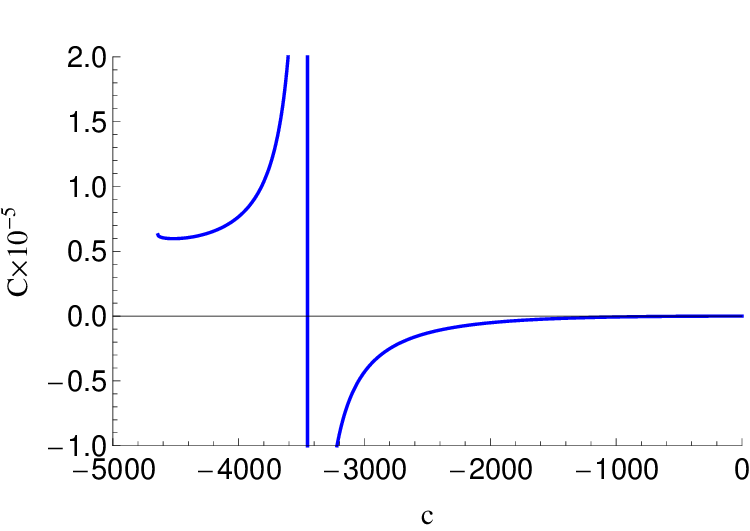} & \epsfxsize=7.5cm %
\epsffile{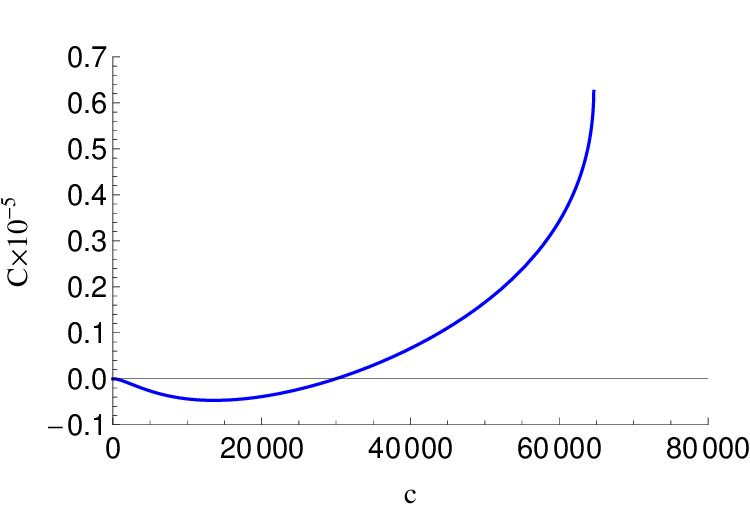} &
\end{array}
$%
\caption{The heat capacity versus $c$ in asymptotically adS spacetime for $%
L=1$ and $r_{+}=100$. Note that the heat capacity is complex valued for $%
c<-4641$ and $c>64641$.}
\label{range}
\end{figure}

In table $V$, we list the QNFs for the fundamental\ mode ($n=0$) and the
first overtone ($n=1$) of intermediate black holes ($r_{+}=1,10$) and large
ones ($r_{+}=50,100$) for $l=0$. From this table, one can see that as the
overtone number and the event horizon radius increase, both the real and
imaginary parts of frequencies increase as well. However, the free parameter
$c$\ affects the frequencies differently based on the size of the event
horizon radius. For intermediate black holes, the QNFs first increase, and
then decrease when $c$ increases. In the case of large black holes, the QNFs
increase linearly with an increase in $c$. Thus, for black holes with $%
r_{+}=100$, the QNMs can be obtained by using the following equations%
\begin{equation}
\begin{array}{cc}
\omega =(184.94213+0.011308c)-i\left( 266.38526+0.00032878c\right) , & \text{%
for }n=0 \\
\omega =(316.12570+0.018958c)-i\left( 491.64179+0.0017483c\right) , & \text{%
for }n=1%
\end{array}%
,  \label{qnmr100}
\end{equation}%
and note that some similar linear equations can be found for different
values of $r_{+}$\ and overtone number. In addition, by considering table $V$%
, we find that the deviation between the QNMs of two consecutive values\ of $%
c$\ decreases when $r_{+}$\ increases, and therefore, changing in $c$\ does
not affect the QNMs significantly in the case of LBHs. Before studying this
behavior and the reason for the linear relation between $\omega $ and $c$,
let us, first, investigate some limitations on Eqs. (\ref{qnmr100}). It is
worthwhile to mention that the equation (\ref{mass}) puts two bounds on the
lower and upper values of $c$. Therefore, $c$ cannot take an arbitrary
negative value to obtain zero $\omega _{r}$\ or $\omega _{i}$. As a result,
there are bounds on the lower and higher values of the QNMs. For example,
for black holes with $r_{+}=100$, $c$\ must obey $-4641\lesssim c\lesssim
64641$ (the heat capacity for this range is illustrated in Fig. \ref{range}%
). Thus, the lower and higher values of the QNMs, based on Eq. (\ref{qnmr100}%
), are given by
\begin{equation}
\begin{array}{cc}
\omega _{\min }\approx 132.462-264.859i;~\omega _{\max }\approx
915.903-287.638i, & \text{for }n=0 \\
\omega _{\min }\approx 228.142-483.528i;~\omega _{\max }\approx
1541.590-604.654i, & \text{for }n=1%
\end{array}%
.
\end{equation}

We should mention that as the imaginary part of QNFs increases, the
corresponds thermal state meets the stability faster. Therefore, the thermal
state with $c\approx 64641$ enjoys the fastest decay rate in its
perturbations. In addition, the Weyl AdS solutions are dynamically stable
under massless scalar perturbations since all the frequencies have a
negative imaginary part, but they are thermally unstable in some areas (see
Fig. \ref{range}).

\begin{center}
\begin{tabular}{ccccccccc}
\hline\hline
$c$ &  & $r_{+}=1$ &  & $r_{+}=10$ &  & $r_{+}=50$ &  & $r_{+}=100$ \\
\hline\hline
$-0.5$ &  & $%
\begin{array}{c}
1.8913-2.9350i \\
3.1966-5.3350i%
\end{array}%
$ &  & $%
\begin{array}{c}
18.4381-26.6371i \\
31.5185-49.1558i%
\end{array}%
$ &  & $%
\begin{array}{c}
92.4598-133.1923i \\
158.0439-245.8191i%
\end{array}%
$ &  & $%
\begin{array}{c}
184.9365-266.3851i \\
316.1162-491.6409i%
\end{array}%
$ \\ \hline
$0$ &  & $%
\begin{array}{c}
2.3588-2.8833i \\
3.9862-5.3126i%
\end{array}%
$ &  & $%
\begin{array}{c}
18.4949-26.6388i \\
31.6136-49.1647i%
\end{array}%
$ &  & $%
\begin{array}{c}
92.4711-133.1926i \\
158.0629-245.8209i%
\end{array}%
$ &  & $%
\begin{array}{c}
184.9421-266.3853i \\
316.1257-491.6418i%
\end{array}%
$ \\ \hline
$0.5$ &  & $%
\begin{array}{c}
2.6297-2.7869i \\
4.4560-5.1969i%
\end{array}%
$ &  & $%
\begin{array}{c}
18.5512-26.6404i \\
31.7081-49.1733i%
\end{array}%
$ &  & $%
\begin{array}{c}
92.4824-133.1930i \\
158.0818-245.8226i%
\end{array}%
$ &  & $%
\begin{array}{c}
184.9478-266.3854i \\
316.1352-491.6427i%
\end{array}%
$ \\ \hline
$1$ &  & $%
\begin{array}{c}
2.7982-2.6712i \\
4.7585-5.0376i%
\end{array}%
$ &  & $%
\begin{array}{c}
18.6070-26.6418i \\
31.8017-49.1816i%
\end{array}%
$ &  & $%
\begin{array}{c}
92.4937-133.1933i \\
158.1008-245.8244i%
\end{array}%
$ &  & $%
\begin{array}{c}
184.9534-266.3856i \\
316.1447-491.6435i%
\end{array}%
$ \\ \hline
$1.5$ &  & $%
\begin{array}{c}
2.8989-2.5446i \\
4.9498-4.8512i%
\end{array}%
$ &  & $%
\begin{array}{c}
18.6624-26.6430i \\
31.8947-49.1896i%
\end{array}%
$ &  & $%
\begin{array}{c}
92.5050-133.1936i \\
158.1197-245.8261i%
\end{array}%
$ &  & $%
\begin{array}{c}
184.9591-266.3858i \\
316.1541-491.6444i%
\end{array}%
$ \\ \hline
$2$ &  & $%
\begin{array}{c}
2.9489-2.4107i \\
5.0575-4.6451i%
\end{array}%
$ &  & $%
\begin{array}{c}
18.7174-26.6441i \\
31.9869-49.1973i%
\end{array}%
$ &  & $%
\begin{array}{c}
92.5163-133.1939i \\
158.1386-245.8279i%
\end{array}%
$ &  & $%
\begin{array}{c}
184.9647-266.3859i \\
316.1636-491.6453i%
\end{array}%
$ \\ \hline
$2.5$ &  & $%
\begin{array}{c}
2.9575-2.2712i \\
5.0575-4.6451i%
\end{array}%
$ &  & $%
\begin{array}{c}
18.7719-26.6450i \\
32.0784-49.2047i%
\end{array}%
$ &  & $%
\begin{array}{c}
92.5276-133.1943i \\
158.1576-245.8296i%
\end{array}%
$ &  & $%
\begin{array}{c}
184.9704-266.3861i \\
316.1731-491.6462i%
\end{array}%
$ \\ \hline
\end{tabular}%
\textbf{\\[0pt]
}

Table $V$: The fundamental mode (first line) and the first overtone (second
line) of the QNFs for $l=0$\ and different values of $c$ and $r_{+} $. The
fundamental frequency and the first overtone of the Schwarzschild black
holes ($c=1$) agree with previous results \cite{Horowitz,SchwADS} calculated
by using Horowitz-Hubeny method.
\end{center}

\begin{figure}[tbp]
$%
\begin{array}{ccc}
\epsfxsize=7.5cm \epsffile{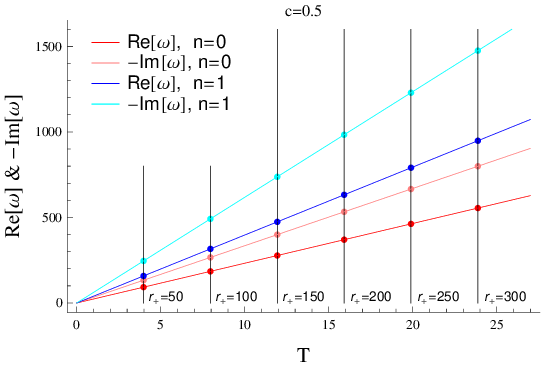} & \epsfxsize=7.5cm \epsffile{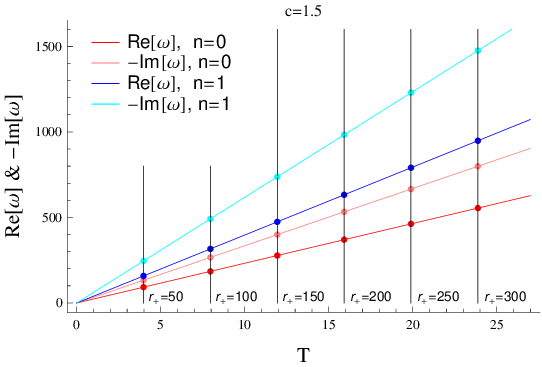} &
\end{array}
$%
\caption{ The QNFs versus temperature for the fundamental mode and first
overtone for $l=0$, $c=0.5$ and $c=1.5$. The vertical lines show the size of
black holes.}
\label{QNMs}
\end{figure}

In order to explain the linear relation between $\omega $ and $c$, and also,
the weak effect of $c$\ on the QNMs of LBHs,\ we consider the temperature (%
\ref{TEMP}) for large black holes%
\begin{equation}
T=\frac{1}{4\pi }\left( \frac{r_{+}}{L^{2}}-\frac{c}{r_{+}}\right) +\mathcal{%
O}\left( \frac{1}{r_{+}^{3}}\right) ,  \label{qnmT}
\end{equation}%
and then look at the relation between the QNMs and this temperature
illustrated in Fig. \ref{QNMs}. As one can see, both the real and imaginary
parts of frequencies increase linearly with an increase in the temperature (%
\ref{qnmT}). On the other hand, for fixed $r_{+}$, the temperature (\ref%
{qnmT}) is a linear function of $c$. Therefore, we can expect a linear
relation between $\omega $ and $c$, as was obtained in (\ref{qnmr100})\ for $%
r_{+}=100$. More interestingly, since $r_{+}$ is present in the denominator
of the second term in (\ref{qnmT}), changing in $c$\ does not affect the
QNMs significantly when $r_{+}$\ increases. Thus, the free parameter $c$\
has a weak effect on the QNMs of LBHs. This behavior was also found
numerically from table $V$.

The points in Fig. \ref{QNMs}, representing the QNMs, lie on straight lines
that their real part are given by%
\begin{equation}
\left\{
\begin{array}{cc}
\omega _{r}=0.0261128+23.2396T, & n=0 \\
\omega _{r}=0.0427779+39.7233T, & n=1%
\end{array}%
\right. \text{ for }c=0.5,  \label{Rc5}
\end{equation}%
\begin{equation}
\left\{
\begin{array}{cc}
\omega _{r}=0.0829902+23.2372T, & n=0 \\
\omega _{r}=0.139923+39.7200T, & n=1%
\end{array}%
\right. \text{ for }c=1.5,  \label{Rc15}
\end{equation}%
while for the imaginary part, we have%
\begin{equation}
\left\{
\begin{array}{cc}
\omega _{i}=0.0264994+33.4738T, & n=0 \\
\omega _{i}=0.0454871+61.7795T, & n=1%
\end{array}%
\right. \text{ for }c=0.5,  \label{Ic5}
\end{equation}%
\begin{equation}
\left\{
\begin{array}{cc}
\omega _{i}=0.0742757+33.4720T, & n=0 \\
\omega _{i}=0.141111+61.7759T, & n=1%
\end{array}%
\right. \text{ for }c=1.5.  \label{Ic15}
\end{equation}

According to the AdS/CFT correspondence, $\tau =1/\omega _{i}$\ is the time
scale of a thermal state to approach thermal equilibrium. Therefore, Eqs. (%
\ref{qnmr100}), (\ref{Ic5}), and (\ref{Ic15}) give the value of $\tau $ and
these equations are the main results of this section. In addition, the QNFs
are linear functions of $r_{+}$\ since the temperature of large black holes
is a linear function of $r_{+}$ (\ref{qnmT}). Interestingly, some similar
results were found for the Schwarzschild-AdS black hole \cite{Horowitz} and
AdS black holes in Born-Infeld massive gravity \cite{BIM}. However, there is
some differences between the Schwarzschild-AdS black hole and Weyl
solutions; for the Schwarzschild case, the QNMs lie on straight lines
through the origin \cite{Horowitz} whereas this is not the case for Weyl
black holes. Besides, the QNMs increase when the new parameter $c$ increases
and the effect of $c$ on the QNMs decrease as $r_{+}$\ increases.

As the final remark, we should mention that by considering the results of
this section and previous results \cite{Horowitz,BIM,Momennia}, the
temperature has a significant role in determining the QNMs of large black
holes in AdS spacetime and nearly extreme black holes in dS/flat spacetimes.
In all cases, the QNM spectrum is a linear function of temperature (surface
gravity for near-extremal black holes).

\section{Conclusions \label{Conclusions}}

We have studied the thermal stability of conformal Weyl black holes in the
canonical ensemble. We observed that the presence of an additional linear $r$%
-term in the Weyl solutions modifies the stability conditions in comparison
with the Schwarzschild-AdS black hole and this linear $r$-term changed the
stability conditions significantly. Although the SBHs of Schwarzschild-(A)dS
solutions are unconditionally unstable for all values of the cosmological
constant, it was possible to find stable SBHs in the conformal Weyl gravity.
In addition, we know that the LBHs of Schwarzschild-(A)dS solutions are
always stable whereas we could find unstable LBHs in Weyl gravity.

Furthermore, we have considered a minimally coupled massive scalar
perturbation in the background spacetime of the conformal Weyl black holes
and calculated the QNFs by using the sixth order WKB approximation and the
AIM after $15$ iterations. We have shown that although the WKB approximation
does not give reliable frequencies for $n\geq l$, this approximation gives
better results for higher $c$ when $n=0=l$. It was shown that most of the
results of WKB approximation are in good agreement with AIM and results get
better for the higher multipole number. Besides, we observed that there were
more oscillations and the modes lived longer for lower $c$. We have shown
that although it is possible to find dynamically stable black holes in the
conformal Weyl gravity, they can be thermally unstable.

In addition, we argued that the black holes have no quasi-resonant
oscillations even when one sets the integration constants $\Lambda =0$\ and $%
c=1$. This happens due to the presence of the linear $r$-term. Therefore,
the imaginary part of the frequencies never vanishes and there are always
damping modes (QNFs are always complex). Moreover, there was a critical mass
for the scalar field such that the decay timescales of the QNMs of the
Schwarzschild-dS black holes and Weyl solutions either grow or decay with
increasing multipole number $l$, depending on whether the mass of the scalar
field was small or large. We have numerically shown that the free parameter $%
c$ affects the value of the critical mass and the critical mass increases
with increasing in $c$.

Also, the QNMs of the Weyl solutions in asymptotically AdS
spacetime were obtained and deviations from those of the
Schwarzschild-adS black holes investigated. The QNFs were
calculated by using the pseudospectral method to investigate the
dynamical stability of the black holes, the effects of the free
parameter $c$\ on the QNMs, and obtain the time scale of the
thermal state to approach thermal equilibrium in conformal field
theory. It was seen that the Weyl solutions are dynamically
stable. Besides, it was shown that the free parameter $c$ affects
the QNMs of the Weyl solutions differently based on the size of
the black hole. As $c$ increased, the QNFs of intermediate black
holes first increased and then decreased. However, the QNMs of
LBHs increased linearly with an increase of $c$. In addition, we
have found that the QNMs are a linear function of the free
parameter $c$ and the event horizon radius (or temperature). We
also have shown that as the size of the black hole increases, the
effect of $c$\ on the QNMs decreases, and thus the effect of $c$\
on the QNMs is negligible in the case of very large black holes.
Since a static large black hole in AdS spacetime corresponds to an
approximately thermal state in CFT, the free parameter $c$ has no
effect on the time scale of the thermal state in the case of very
large black holes. As a result, two thermal states correspond to
extremely large black holes in Einstein and Weyl gravities are
identical.

\begin{acknowledgements}
We wish to thank Shiraz University Research Council. MM wishes to
thank A. Zhidenko and R. A. Konoplya for their helps on QNMs.
\end{acknowledgements}

\end{document}